\documentclass{aastex631}

\usepackage{amsmath}
\usepackage{tasks}
\usepackage{booktabs}
\usepackage{rotating}
\usepackage{makecell}
\usepackage{xspace}
\usepackage{enumitem}
\usepackage{textcomp}
\usepackage{stackengine}
\setlist[itemize]{noitemsep}

\shorttitle{The Carousel Lens: A Well-Modeled Strong Lens with Multiple Lensed Sources}
\shortauthors{Sheu et al.}

\graphicspath{{./}{figures/}}

\newcommand{\ipg}{\emph{g}\xspace}
\newcommand{\ipr}{\emph{r}\xspace}
\newcommand{\ipi}{\emph{i}\xspace}
\newcommand{\ipz}{\emph{z}\xspace}

\newcommand{\txr}{\textcolor{black}}
\newcommand{\txo}{\textcolor{black}}

\newcommand{\sn}[2]{\ensuremath{{#1}\times 10^{#2}}}
\newcommand\angdot[1][\circ]{%
  \stackengine{0pt}{.}{${}^{\mathrm{#1}}$}{O}{l}{F}{F}{L}}
\newcommand\angdotcustom[1][\circ]{%
  \stackengine{0pt}{.}{${\mathrm{#1}}$}{O}{l}{F}{F}{L}}

\begin{document}

\title{The Carousel Lens: A Well-Modeled Strong Lens with Multiple Sources
%5+ Lensed Sources,
Spectroscopically Confirmed by VLT/MUSE}

\correspondingauthor{William Sheu}
\email{wsheu@astro.ucla.edu}

\author[0000-0003-1889-0227]{William Sheu}
\affiliation{Department of Physics \& Astronomy, University of California, Los Angeles, 430 Portola Plaza, Los Angeles, CA 90095, USA}

\author[0000-0001-7101-9831]{Aleksandar Cikota}
\affiliation{Gemini Observatory / NSF's NOIRLab, Casilla 603, La Serena, Chile}

\author[0000-0001-8156-0330]{Xiaosheng Huang}
\affiliation{Department of Physics \& Astronomy, University of San Francisco, 2130 Fulton Street, San Francisco, CA 94117-1080, USA}
\affiliation{Physics Division, Lawrence Berkeley National Laboratory, 1 Cyclotron Road, Berkeley, CA 94720, USA}

\author[0000-0002-3254-9044]{Karl Glazebrook}
\affiliation{Centre for Astrophysics and Supercomputing, Swinburne University of Technology, Hawthorn, Victoria 3122, Australia}
\affiliation{ARC Centre of Excellence for All Sky Astrophysics in 3 Dimensions (ASTRO 3D), Australia}

\author[0000-0002-0385-0014]{Christopher Storfer}
\affiliation{Physics Division, Lawrence Berkeley National Laboratory, 1 Cyclotron Road, Berkeley, CA 94720, USA}
\affiliation{Institute for Astronomy, University of Hawaii, 2680 Woodlawn Drive, Honolulu, HI 96822-1897, USA}

\author[0000-0002-2350-4610]{Shrihan Agarwal}
\affiliation{Department of Astronomy, University of Chicago, 5640 S Ellis Ave, Chicago, IL 60615, USA}

\author[0000-0002-5042-5088]{David J. Schlegel}
\affiliation{Physics Division, Lawrence Berkeley National Laboratory, 1 Cyclotron Road, Berkeley, CA 94720, USA}

\author[0000-0001-7266-930X]{Nao Suzuki}
\affiliation{Physics Division, Lawrence Berkeley National Laboratory, 1 Cyclotron Road, Berkeley, CA 94720, USA}
\affiliation{Department of Physics, Florida State University, 77 Chieftan Way, Tallahassee, FL 32306, USA}
\affiliation{Kavli Institute for the Physics and Mathematics of the Universe, University of Tokyo, Kashiwa 277-8583, Japan}

\author[0000-0002-2784-564X]{Tania M. Barone}
\affiliation{Centre for Astrophysics and Supercomputing, Swinburne University of Technology, Hawthorn, Victoria 3122, Australia}
\affiliation{ARC Centre of Excellence for All Sky Astrophysics in 3 Dimensions (ASTRO 3D), Australia}

\author[0000-0002-1620-0897]{Fuyan Bian}
\affiliation{European Southern Observatory, Alonso de C\'{o}rdova 3107, Casilla 19001, Vitacura, Santiago 19, Chile}

\author[0000-0001-6089-0365]{Tesla Jeltema}
\affiliation{Santa Cruz Institute for Particle Physics, Santa Cruz, CA 95064, USA}

\author[0000-0001-5860-3419]{Tucker Jones}
\affiliation{Department of Physics and Astronomy, University of California, Davis, 1 Shields Avenue, Davis, CA 95616, USA}

\author[0000-0003-1362-9302]{Glenn G. Kacprzak}
\affiliation{Centre for Astrophysics and Supercomputing, Swinburne University of Technology, Hawthorn, Victoria 3122, Australia}
\affiliation{ARC Centre of Excellence for All Sky Astrophysics in 3 Dimensions (ASTRO 3D), Australia}

\author[0000-0003-4083-1530]{Jackson H. O'Donnell}
\affiliation{Santa Cruz Institute for Particle Physics, Santa Cruz, CA 95064, USA}

\author[0000-0002-2645-679X]{Keerthi Vasan G. C.}
\affiliation{Department of Physics and Astronomy, University of California, Davis, 1 Shields Avenue, Davis, CA 95616, USA}

\begin{abstract}

Over the past few years alone, the lensing community has discovered thousands of strong lens candidates, and spectroscopically confirmed hundreds of them.  In this time of abundance, it becomes pragmatic to focus our time and resources on the few extraordinary systems, in order to most efficiently study the universe.  In this paper, we present such a system: DESI-090.9854-35.9683, a cluster-scale lens at $z_{\rm l} = 0.49$, 
with seven observed lensed sources around the core, and additional lensed sources further out in the cluster.  From the number and the textbook configuration of the lensed images, a tight constraint on the mass potential of the lens is possible.  This would allow for detailed analysis on the dark and luminous matter content within galaxy clusters, as well as a probe into dark energy and high-redshift galaxies.  We present our spatially resolved kinematic measurements of this system from the Very Large Telescope Multi Unit Spectroscopic Explorer, which confirm five of these source galaxies (in ascending order, at $z_{\rm s} = 0.962, 0.962, 1.166, 1.432, \text{ and } 1.432$). 
With previous \textit{Hubble Space Telescope} imaging in the F140W and F200LP bands, we also present a simple two power-law profile flux-based lens model that, for a cluster lens, well models the five lensed arc families with redshifts.
We determine the mass to be $M(< \theta_{\rm E}) = 4.78\times10^{13} M_{\odot}$ for the primary mass potential.
From the model, we extrapolate the redshift of one of the two source galaxies not yet spectroscopically confirmed to be at $z_{\rm s}=4.52^{+1.03}_{-0.71}$.  

\end{abstract}

\keywords{Strong gravitational lensing --- Galaxy clusters --- Galaxy spectroscopy --- High-redshift galaxies}

\section{Introduction} \label{sec:introduction}
\begin{figure*}
\begin{center}
 \includegraphics[width=1\linewidth]{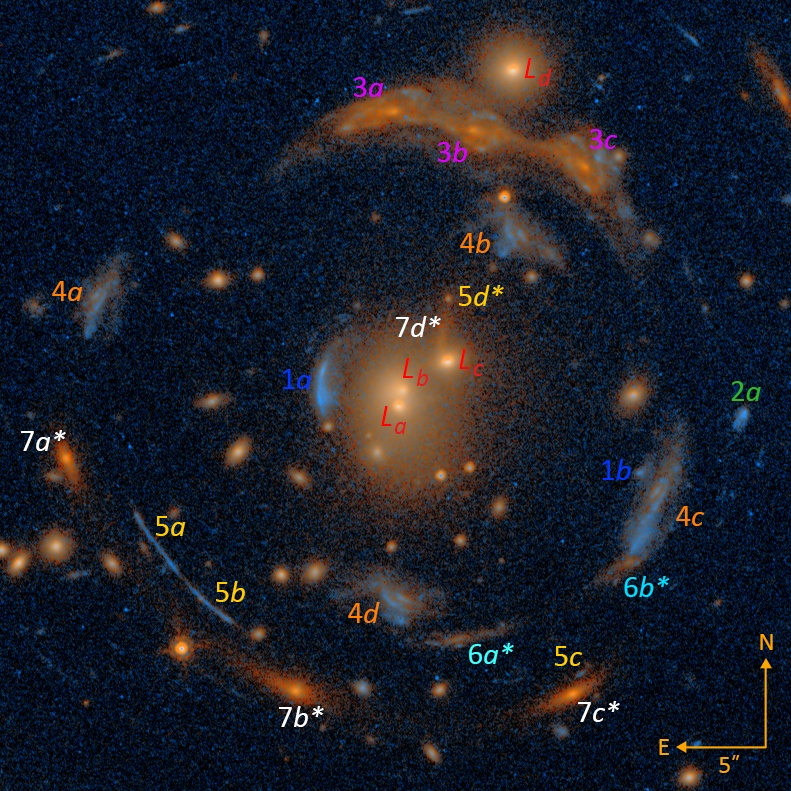}
 \caption{RGB image of DESI-090.9854-35.9683, generated using the \textit{HST} F140W filter \txr{(0.070$''$/pixel)} as the red channel, \textit{HST} F200LP filter \txr{(0.028$''$/pixel)} as the blue channel, and the pixel-wise average of the two filters as the green channel.  The brightest cluster-member galaxies within our cutout, labeled as $L_a$, $L_b$, $L_c$, and $L_d$, are approximately at the same redshift of $z_{\rm l} = 0.49$ (see Table~\ref{tab:lens_redshifts}).  The source galaxy images, labeled with a number prefix and a letter, are organized such that images of the same numeral prefix and color correspond to the same source galaxy.  
 Note that $2a$ is the only image for this arc family. 
 Image labels followed by an asterisk indicate that they do not exhibit prominent lines in the MUSE IFU data, and hence we cannot spectroscopically confirm their redshifts.
 The spectroscopic redshifts of galaxies 1, 2, 3, 4, and 5 are $z_{\rm s} = 0.962, 0.962, 1.166, 1.432, \text{ and } 1.432$, respectively (see \S\ref{sec:muse}).}
 \label{fig:rgb}
\end{center}
\end{figure*}

Strong lensing occurs when the gravity of a massive object, typically a galaxy or galaxy cluster, distorts the light path from a background source galaxy
with near perfect alignment with the observer.
% within our line-of-sight.  
These distortions can result in multiple, magnified, and warped images of the source galaxy surrounding the lensing object.  
% When properly analyzed, 
These systems are invaluable probes to many facets of cosmology.  By fitting for the lensed images and analyzing the lens galaxy kinematics, we can study the 
% observationally-invisible 
dark matter profiles of the lens \citep[e.g.,][]{shajib2021}.  By measuring and modeling the time delays of lensed transient events or variable objects between the multiple images, we can constrain the Hubble constant $H_0$, independent of the CMB and direct distance ladder measurements, which currently disagree \citep[e.g.,][]{wong2020}.  
By virtue of its ability to drastically magnify background sources, strong lensing  allows for significantly deeper and further observations than otherwise possible
% what was originally indented for the observing instrument 
\citep[e.g.,][]{roberts2022}. 

This \txr{paper} is organized as follows: we describe the general properties and discovery status of DESI-090.9854-35.9683 (\S\ref{sec:desi090}), 
provide our analysis of the Multi Unit Spectroscopic Explorer (MUSE) integral field units (IFU) data (\S\ref{sec:muse}), present our lens model of the system (\S\ref{sec:modeling}), 
discuss the possible redshift of one of the lensed sources (\S\ref{sec:redshiftarc}),
and conclude (\S\ref{sec:conclusion}).

\section{DESI-090.9854-35.9683 Discovery and Properties} \label{sec:desi090}
DESI-090.9854-35.9683 (RA: $6^{\rm h} 3^{\rm m} 56 \angdot[{\rm s}] 50$, dec.: $-35^{\circ} 58' 5\angdotcustom[{''}] 88$) was first identified in \citet{jacobs2019} as a ``possible, but not probable or definite'' strong lens system candidate by applying a convolutional neural network to the Dark Energy Survey (DES) Year 3 imaging.  Its low score is likely due to the small cutouts ($26'' \times 26''$) used during visual inspection, which missed the large cluster-scale lensing nature \citep{jacobs2019}.
\citet{huang2021} later independently identified this system with a residual neural network on the Dark Energy Spectroscopic Instrument (DESI) Legacy Imaging Surveys Data Release 8; following visual human inspection, the system was given a strong lens candidacy grade of A (see their Figure~10).  
This system was also found by \citet{odonnell2022a} as DESJ0603-3558, by applying a color-magnitude selection to catalogs created from the first three observing seasons of DES, followed by visual scanning.
DESI-090.9854-35.9683 was observed by the \textit{Hubble Space Telescope} (\textit{HST}) Wide Field Camera 3 for 600 seconds each in the F140W and F200LP filters (\textit{HST} proposal \#16773; K. Glazebrook), which is shown as an RGB image in Figure~\ref{fig:rgb}. 
All the \textit{HST} data used in this paper can be \txr{obtained from the MAST archive at} \dataset[10.17909/zq07-4f53]{http://dx.doi.org/10.17909/zq07-4f53}.  
Throughout this \txr{paper}, we will reference the labeling scheme illustrated in Figure~\ref{fig:rgb}.  

Strong lensing in DESI-090.9854-35.9683 occurs around the center of a small cluster with \txr{four} spectroscopically confirmed source galaxies, \txr{plus one weakly lensed galaxy}.  While this is a cluster lens, we note that the lens profile does not seem overly complex, as the lensed image configurations for arc families 1, 3, 4, and 7 are typical of simple lens profiles, implying a fairly relaxed lensing potential.  Source galaxy 1 ($z_{\rm s} = 0.962$; see \S\ref{sec:muse}) is doubly lensed.  %the most massive and brightest member galaxies (with the exception of $L_d$) appear to be close to the potential center and
Source galaxy 2 ($z_{\rm s} = 0.962$) appears to have only one lensed image\txr{, and hence is weakly lensed}.  
The lensed images of source galaxy 3 ($z_{\rm s} = 1.166$) appear to be in a ``naked cusp''  configuration \citep[e.g.,][]{brada2004}, given that there are three bright images merging, and yet a fourth image is not present in the MUSE data cube nor in our model predictions.  Such a configuration is only possible if the lens' inner diamond caustic extends outside its elliptical caustic, which gives additional insight on the lens potential \citep[e.g.,][]{narayan1992, kochanek1992}.  Also, the source galaxy 3 images are significantly perturbed by one of the cluster member galaxies $L_d$, in addition to the main cluster profile centered near galaxies $L_a$, $L_b$, and $L_c$.  
Source galaxy 4 ($z_{\rm s} = 1.432$) is quadruply lensed in an Einstein cross configuration; this source exhibits a high level of structural complexity.  
Source galaxy 5 ($z_{\rm s} = 1.432$) appears to be quadruply lensed, with a demagnified fourth image $5d$ predicted by our model (see \S\ref{sec:modeling}).  
% Not much is known about 
Given the similarity in their colors, morphology, and their tangential elongation, we posit that objects $6a$ and $6b$ are lensed images of the same source galaxy,  
% due to their similarity in color and shape, 
but no spectroscopic redshift can be identified from the MUSE data (though the MUSE data cube does indeed distinguish object $6b$ from $4c$).  
Source galaxy 7 is quadruply lensed, with $7d$ being a radial arc.
Not only is the color of $7d$ consistent with $7a - 7c$, but the location is in agreement with lens modeling (see \S\ref{sec:redshiftarc}).  
There are no prominent spectroscopic lines present for source galaxy 7 in the MUSE observations, but our lens modeling predicts a redshift of \txr{$z_{\rm s}=4.52^{+1.03}_{-0.71}$} (also in \S\ref{sec:redshiftarc}), which would make it an interesting probe into high-redshift galaxies.

In addition to its prominent lensing features, DESI-090.9854-35.9683 was also identified as a Sunyaev-Zel’dovich (SZ) galaxy cluster through X-ray detections in the ROSAT All-Sky Survey \citep{voges2000, flesch2010, tarrio2019}, and microwave detections from the Atacama Cosmology Telescope \citep{hilton2021}.

\section{MUSE IFU Observations} \label{sec:muse}
DESI-090.9854-35.9683 was observed on September 30, 2023 at 07:45h UT %as part of an European Southern Observatory (ESO) filler program for characterizing of 
% galaxy–galaxy 
%strong gravitational lensing system candidates 
(Prog. ID 0111.B-0400(H)) with MUSE \citep{Bacon_2010}, an IFU spectrograph mounted at UT4 of ESO's Very Large Telescope (VLT) on Cerro Paranal in Chile.  The observations were carried out in the MUSE WFM-NOAO-N mode, with a field of view of $60'' \times 60''$ and a %spatial resolution 
spaxel size of 0.2$''$.  The spectral range is from 4750 to 9350~\AA\ with the spectral resolution ranging from R = 2000 to 4000~\AA\ across the wavelength domain.
The observations were taken with $4 \times 700$ second exposures during DIMM seeing of $\sim$0.9$''$ in clear but windy ($\sim$10 m/s) weather, and reduced following standard procedures with the MUSE pipeline package version 2.2 \citep{2020A&A...641A..28W} that is a part of the ESO Recipe Execution Tool (ESOREX). We also removed sky lines using the Zurich Atmosphere Purge (ZAP) sky subtraction tool \citep{2016MNRAS.458.3210S}.
\begin{deluxetable}{cccccccc} \label{tab:lens_redshifts}
\tablewidth{0pt} 
\tablecaption{Lens and source redshifts, and their statistical uncertainties. The redshifts have been determined by fitting the Ca H\&K line in the case of the $L_{a+b}$, $L_c$ and $L_d$, or the [\ion{O}{2}] doublet in the case of the lensed images of the sources ($1a - 5b$). The proper velocity difference is calculated as $v_i = c(z_i-z)/(1+z)$, where $z$ is the average of the images redshifts for each source (or the average of the cluster-member galaxy redshifts in the case of $L_{a+b}$, $L_b$, and $L_c$).  The proper velocity for $2a$ is calculated assuming it is of the same source as $1a$ and $1b$, resulting in the outlying velocity presented.}

\tablehead{
\colhead{Object} & \colhead{Redshift} & \colhead{Proper velocity difference} \\
& &  [km s$^{-1}$]  
}
\startdata
$L_{a+b}$ & $0.48784 \pm 0.00045$ & $-123.5 \pm 98.8$ \\
$L_c$ & $0.48716 \pm 0.00033$ & $-260.5 \pm 77.4$ \\
$L_d$ & $0.49036 \pm 0.00019$  &  $384.0 \pm 55.1$ \\ \\
$1a$  & $0.96196 \pm 0.00001$ & $25.0 \pm 4.2$ \\
$1b$ & $0.96189 \pm 0.00007 $ &  $14.3 \pm11.4$ \\ \\
$2a$ & $0.96154 \pm 0.00003 $ & $-39.2 \pm 6.0$ \\ \\
$3a$  & $1.1657 \pm 0.0001$  & $-38.6 \pm 17.9$\\
$3b$  & $1.1659 \pm 0.0002$  & $-19.4 \pm 29.9$\\
$3c$ & $1.1664 \pm 0.0001$  & $58.1 \pm 17.8$ \\ \\
$4a$  & $1.43227 \pm 0.00003 $ & $-4.9 \pm 4.0$\\
$4b$  & $1.43230 \pm 0.00002 $ & $-1.2 \pm 2.8$\\
$4c$  & $1.43233 \pm 0.00002 $ & $2.5 \pm 2.8$\\
$4d$  & $1.43234 \pm 0.00002$ &  $3.7 \pm 2.8$\\ \\
$5a$ & $1.43198 \pm 0.00004 $ & $-1.9 \pm 5.8$\\
$5b$  & $1.43201 \pm 0.00003$ & $1.9 \pm 4.8$ \\
\enddata
\end{deluxetable}

\begin{figure*}
\begin{center}
\includegraphics[width=1\linewidth]{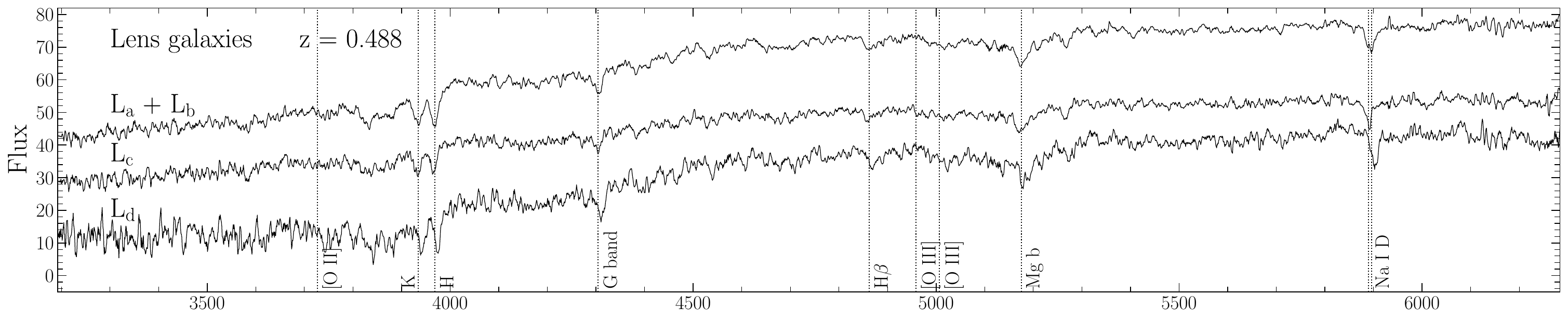}
\includegraphics[width=1\linewidth]{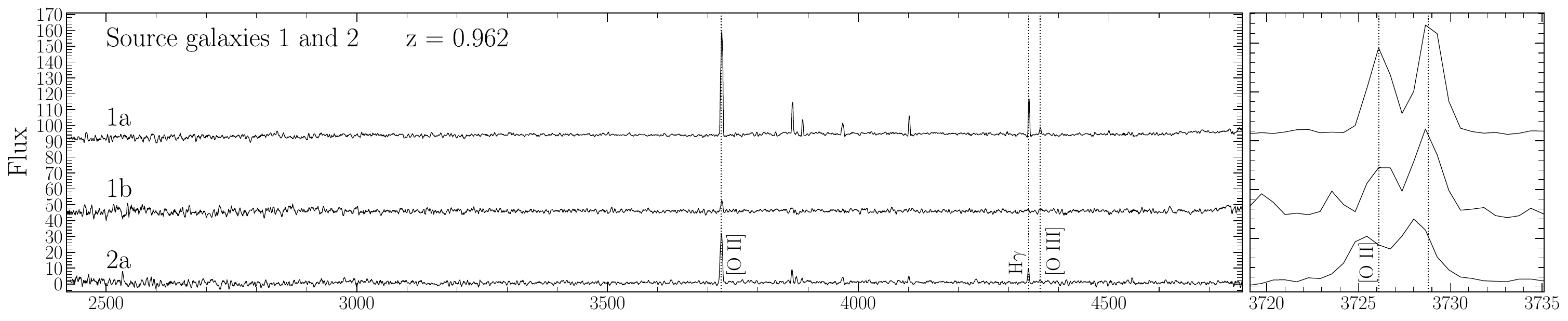}
\includegraphics[width=1\linewidth]{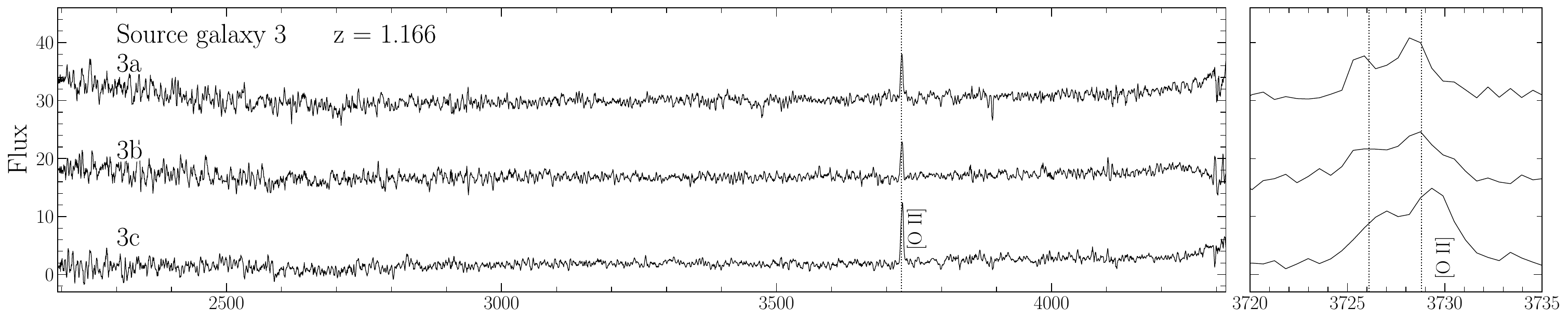}
\includegraphics[width=1\linewidth]{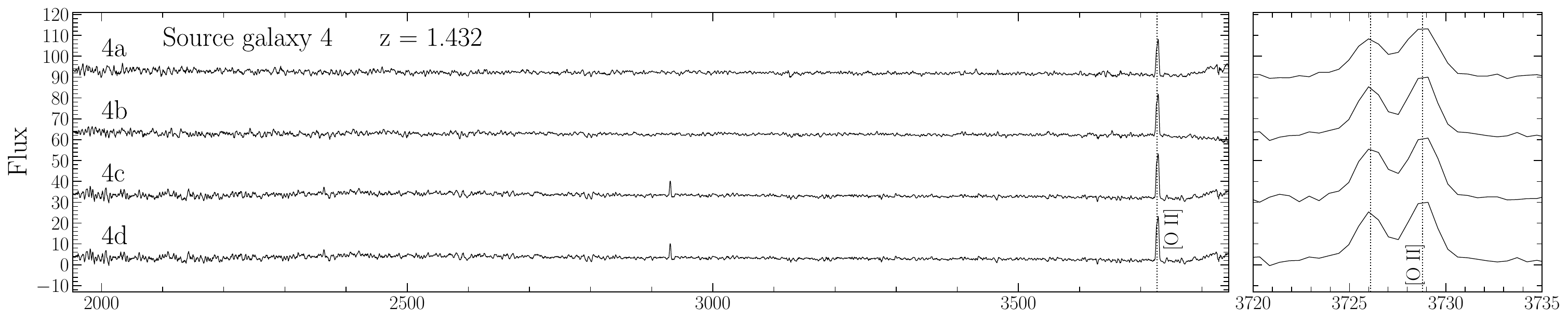}
\includegraphics[width=1\linewidth]{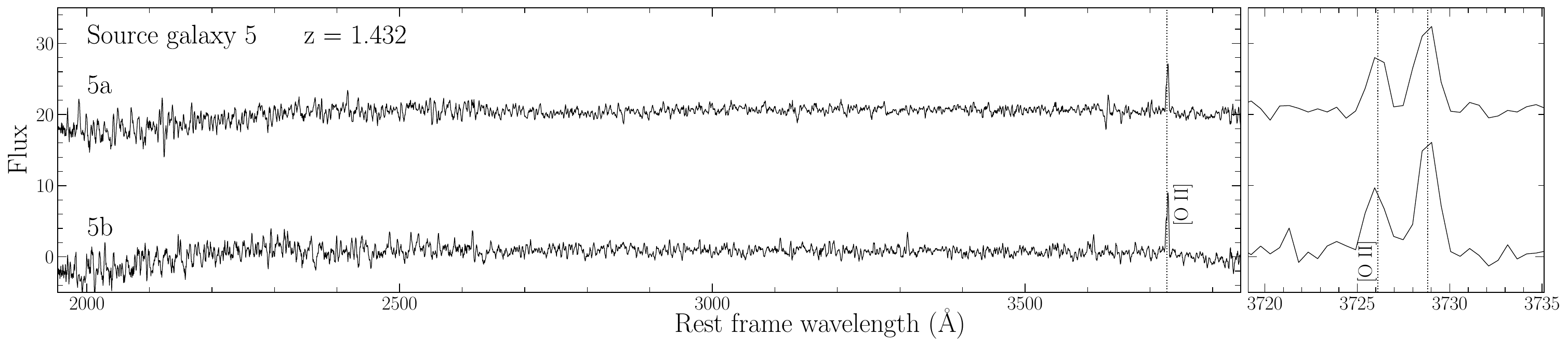}
 \caption{Spectra of the lens and images of the sources extracted from the MUSE cube. The spectra are the average of multiple hand-selected spaxels for each object (see Figure~\ref{fig:MUSE_velocities}). The right panels for the sources are cut-outs around the [\ion{O}{2}] $\lambda$$\lambda$ 3726.1, 3728.8 doublet line used for the redshift determination.}
 \label{fig:MUSE_spectra}
\end{center}
\end{figure*}

We extracted the spectra of the lens and the lensed images of the sources from the MUSE data cube and determined the redshifts by matching prominent emission and absorption lines to the spectra (Figure~\ref{fig:MUSE_spectra}). Using the Ca H\&K lines, we successfully determine the redshifts of the lensing galaxies ($L_a + L_b$, see Figure~\ref{fig:rgb} for a map of objects), which also display prominent G band, Mg b, and \ion{Na}{1} D absorption lines, and of the lensed images of source galaxies 1, 2, 3, 4, and 5 on account of the [\ion{O}{2}] $\lambda \lambda$ 3726.1, 3728.8 doublet. All redshifts are given in Table~\ref{tab:lens_redshifts}.
The lensing galaxies, $L_{a+b}$, are part of a galaxy cluster. The other two bright members, $L_c$ and $L_d$, also display prominent absorption lines at a redshift consistent with proper velocities within galaxy clusters.  From these three measurements, we calculate a mean cluster redshift of $z = 0.4884 \pm 0.0014$.  We also note that the source galaxies 1 and 2 have similar redshifts of $z_{\rm s} = 0.962$, and source galaxies 4 and 5 have similar redshifts of $z_{\rm s} = 1.432$. This may indicate that we are observing a series of three galaxy groups/clusters that lie on the same line-of-sight. Lastly, source galaxy 3 has a redshift of $z_{\rm s} = 1.166$. %, and a mean proper absolute velocity of $256 \pm 135 \text{ km s}^{-1}$.   

\begin{figure*}
\begin{center}
\includegraphics[height=3 in]{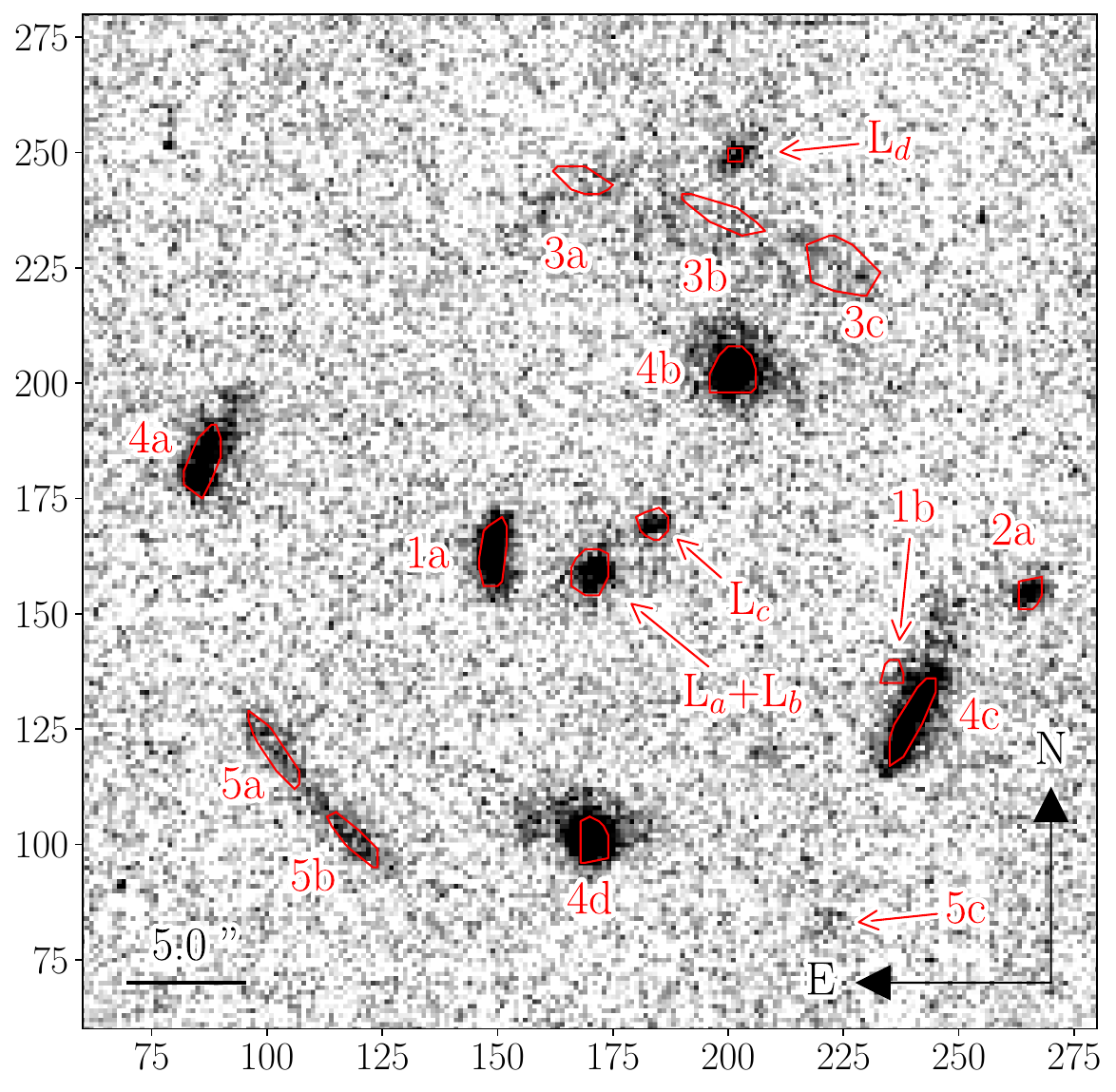}
\includegraphics[height=3 in]{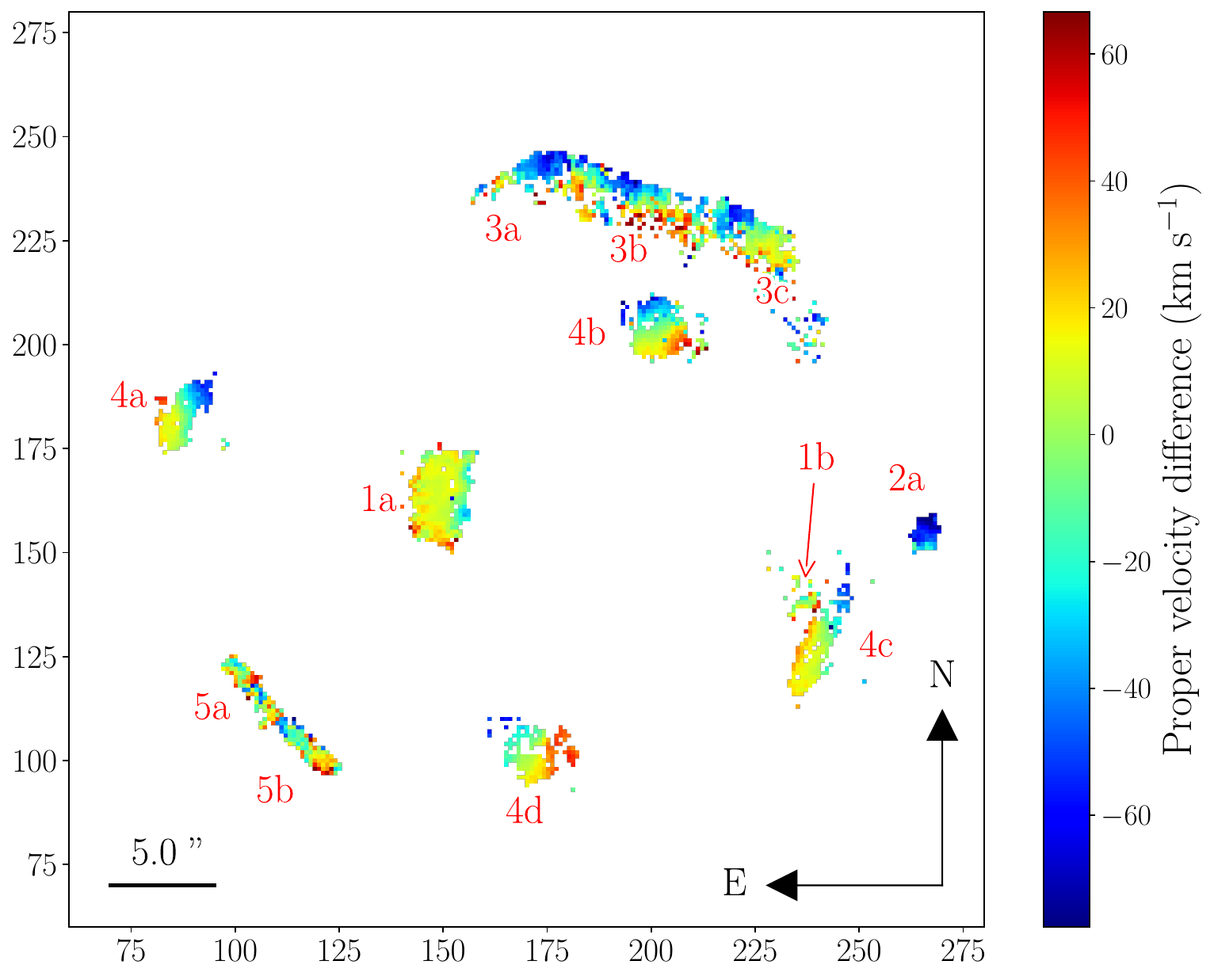}
 \caption{Cutouts of the MUSE data of DESI-090.9854-35.9683. The scale is in pixels, where the pixel size is $0.2''$. North is up, east left. Left: the [\ion{O}{2}]$\lambda$$\lambda$ 3726.1, 3728.8 line-map of the images 1, 2, 3, 4, and 5.  The red outlines illustrate the spaxels used in generating the spectra in Figure~\ref{fig:MUSE_spectra}.  Right: the proper velocity map of the lensed images. The velocity difference is relative to the average redshift for each respective source.} %, and the collapsed spectrum of the lens
\label{fig:MUSE_velocities}
\end{center}
\end{figure*}

The lensed galaxy 1 displays 2 arcs, $1a$ and $1b$ at consistent redshifts (see Figure~\ref{fig:MUSE_spectra}). Initially, the bright image 2 was thought to be the counterarc of $1a$; however, lens modeling (see \S\ref{sec:modeling}) showed inconsistencies with the expected position of the image $1b$. Furthermore, the redshift of $2a$ is also inconsistent with the redshift of $1a$ (with a $6.3 \sigma$ discrepancy; see Figure~\ref{fig:MUSE_spectra} and Table~\ref{tab:lens_redshifts}).
Figure~\ref{fig:MUSE_velocities} shows a proper velocity difference map of the images, and it is clear that, while all the other images have consistent velocities, the velocity of source 2 is an outlier.
A closer inspection of the MUSE cube reveled a faint counterarc, $1b$, at the expected position. We conclude that image $2a$ belongs to a singly lensed galaxy distinct from arc family 1.  In image $1a$, the [\ion{O}{3}] is detected, which could be used to probe the gas-phase metallicity of source galaxy 1 in future analysis.

Source galaxy 3 displays three lensed images north of the lens: $3a$, $3b$, and $3c$, at a redshift of $z_{\rm s} = 1.166$. 
In Table~\ref{tab:lens_redshifts}, we see significant differences in the proper velocity between images $3a$, $3b$, and $3c$, which we associate as being due to their diffused imaging in the MUSE cube (see Figure~\ref{fig:MUSE_velocities}).  Because of this, we cannot well ascertain which part of the source galaxy we are probing through the spectra of each image.  Therefore, the significant difference in proper velocities are likely due to rotation/dispersion within the source galaxy, and our spaxel selection capturing different regions of the source galaxy for each image.     %.  That is, the difference in proper velocities are likely due to the images blending into each other (and hence uncertainty in the image locations) within the MUSE data.  There is a difference in the parity between the images visible in the velocity map (Figure~\ref{fig:MUSE_velocities}). The left (east) part of the images $3b$ and $3c$ is blueshifted while the right part (west) is redshifted. In contrast, the left (east) part of the image $3a$ is redshifted while the right part (west) is blueshifted. This difference is also consistent with the image parity prediction from the lens model, as images $3a$ and $3b$ are merging across the critical curve passing between them (see \S\ref{sec:modeling}).  

The source galaxies 4 and 5 are both quadruply lensed and at a similar redshift of $z_{\rm s} = 1.432$. 
The quad lensed images of source galaxy 4 exhibit textbook parities (see right panel of Figure~\ref{fig:MUSE_velocities}), indicating that the central part of the cluster is dynamically relaxed.
This, together with the overall relatively high degree of symmetry of this system, motivated a relatively simple lens model in the next section.
Although their redshifts are similar, there is a statistically significant difference between source 4 and 5 (of $6.2 \sigma$; see Table~\ref{tab:lens_redshifts}).
Source galaxy 5 displays two bright arcs, clearly visible in the MUSE data (see left panel in Figure~\ref{fig:MUSE_velocities}), $5a$ and $5b$. The position of image $5c$ is predicted by the lens model and found in the MUSE datacube after closer examination.  The image $5c$ is faint, and so we do not show its spectrum in Figure~\ref{fig:MUSE_spectra}, but the [\ion{O}{2}] emission is visible in the line-map (see left panel in Figure~\ref{fig:MUSE_velocities}).  $5d$ is also predicted by the lens model, but as it is demagnified and therefore even fainter, it cannot be spectroscopically confirmed by the MUSE data.

The images of source galaxy 6 are visible in the Near-IR \textit{HST} images, but very faint in the \emph{HST} optical F200LP filter, and are not visible in the MUSE (optical) data.  The images of source 7 (Figure~\ref{fig:clusterplot}) are faint, but visible in the collapsed MUSE image. However the extracted spectra do not show any prominent lines, and so we can not determine the redshift (see more about source galaxy 7 in \S\ref{sec:redshiftarc}).  
 
In the spectra for source galaxies $1-5$ (Figure~\ref{fig:MUSE_spectra}), the [\ion{O}{2}] doublet is well resolved.  Not only does this allow for accurate redshift measurements (as shown in this \txr{paper}), but it can also be used to probe the densities of these source galaxies.  We leave the detailed spectral analysis for future works.  Though we do not yet have spectroscopic redshifts for arc families 6 and 7 (6$a$, $b$ and 7$a$, $b$, $c$, respectively), based on the \emph{HST} observations (Figure~\ref{fig:rgb}), it is clear that they are lensed arcs:
1) by the arrangement of images in these two arc families, namely, they form an arc with the center of curvature matching the center of the cluster; 
and 2) by the similarity in morphology and color for images 6$a$ and 6$b$, and for images 7$a$, $b$ and $c$, respectively. 
More quantitatively for images 7$a$, $b$, and $c$, 
Table~\ref{tab:mags} shows that they have very similarly SEDs.

\section{Lens Modeling} \label{sec:modeling}
We use \textsc{lenstronomy}\footnote{\url{https://github.com/lenstronomy/lenstronomy}}, a multi-purpose lens modeling software package, to model DESI-090.9854-35.9683, with multiple source redshift planes to account for each source galaxy \citep{birrer2018, lenstronomyII}.  \textsc{lenstronomy} implements a forward-modeling algorithm to estimate the lens potential and surface brightness profiles, by reconstructing an image of the lensing system and minimizing over the pixel-level residuals with the data.  In the Time-Delay
Lens Modeling Challenge \citep[TDLMC;][]{ding2021}, \textsc{lenstronomy} has proved robust and accurate, it was used by multiple teams to recover lens model parameters within statistical consistency (in Rung 2).  

In this \txr{paper}, we model based on the F140W image as opposed to the F200LP, as the SNR in source galaxies 3 and 4 are 
% visibly 
clearly higher in the F140W filter.  
% We implement a simple 
Our model consists of two elliptical power law lens profiles (both at $z_{\rm l}=0.49$), centered at object $L_a$ and $L_d$ respectively.  
This was the best model (compared to other iterations of power law lenses centered at $L_a$, $L_b$, $L_c$, and $L_d$) based on the lowest Bayesian information criterion (BIC) value using solely lensed image positions.  As objects $L_a$, $L_b$, $L_c$, and $L_d$ are the brightest cluster members in the cutout, their light profiles are modeled with elliptical S\'{e}rsic profiles \citep{sersic} so as to prevent their lens light from contaminating the lensed images.  All other interloping objects, as well as source galaxy 6 and 7 images (since we do not have spectroscopic redshifts for these), are masked out (see the top right panel of Figure~\ref{sec:modeling}).  The light profiles for source galaxies 1, 3, 4, and 5 are each modeled with an elliptical S\'{e}rsic profile and a shapelets basis \citep{shapelets1, shapelets2, birrer2015a} of order 10, 15, 9, and 8, respectively. % (at redshifts of 0.96, 1.17, 1.43, and 1.43, respectively).  
These orders are determined through trial and error to improve the model fit with the best reduced $\chi^2$ \citep[e.g.,][]{tan2024}.  A Bayesian model comparison is often used to assess and determine how a given model fits the image data \citep[e.g.,][]{Nightingale2018, Shajib2018}, but given the number of sources and the permutations of orders, this method would be computationally impractical.
For source galaxy 2, we only implement a single elliptical S\'{e}rsic profile, as there is only one image present, and its appearance does not necessitate additional complexity.  \txr{We note that while source 2 (being singly imaged) does not contribute as much constraining power to model as compared to the other lensed sources, the fact that it is singly imaged narrows the lensing parameter space by forcing the delensed image to lie outside the caustic.  While it can be argued that this parameter space is already being excluded by the other image constraints, we nevertheless include source 2 into our model as further confirmation.}  We model the point spread function (PSF) by stacking nearby stars in the \textit{HST} exposure\footnote{\txr{We apply the implementation provided in \url{https://github.com/sibirrer/AstroObjectAnalyser}.}}.
\begin{deluxetable}{cccccccc} \label{tab:lens_model}
\tablewidth{0pt} 
\tablecaption{Lens model parameters and their statistical uncertainties.  Here, $\theta_{\rm E}$ is the Einstein radius (with respect to $z_{\rm s} = 1.432$), $\gamma$ is the the logarithmic slope of the mass profile, $q$ is the mass axis ratio, PA is the mass position angle, $M(< \theta_{\rm E})$ is the projected mass within the Einstein radius, $\gamma^{\text{ext}}$ is the external shear magnitude, and $\phi^{\text{ext}}$ is the external shear angle. 
 All angular measures are given as North of East.}

\tablehead{
\colhead{Object} & \colhead{$\theta_{\rm E}$} & \colhead{$\gamma$} & \colhead{$q$} & \colhead{PA} & \colhead{$M(< \theta_{\rm E})$} & \colhead{$\gamma^{\text{ext}}$} & \colhead{$\phi^{\text{ext}}$}\\
  & [$''$] &  &  & [deg] & [$10^{11} M_\odot$] &  & [deg] }
\startdata
$L_a$ & $13.03 \pm 0.02$ & $1.67 \pm 0.01$ & $0.87 \pm 0.01$ & $-45 \pm 1$ &  $478 \pm 2$ & - & - \\
$L_d$ & $0.99 \pm 0.02$ & $2.12 \pm 0.01$ & $0.69 \pm 0.01$ & $-38 \pm 1$ & $2.77 \pm 0.12$ & - & - \\
External Shear & - & - & - & - & - & $0.11 \pm 0.01$ & $9 \pm 1$ \\
\enddata
\end{deluxetable}

\begin{figure*}
\begin{center}
 \includegraphics[width=1\linewidth]{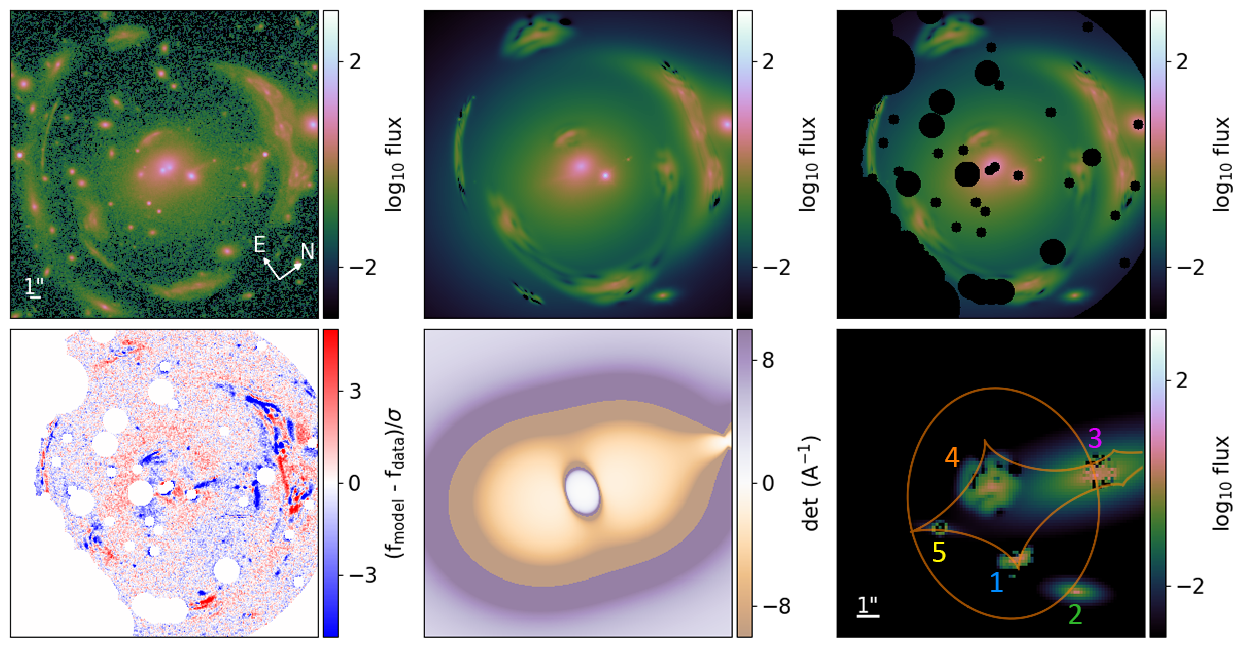}
 \caption{Modeling results for DESI-090.9854-35.9683.  Top left: the \textit{HST} F140W image.  Top middle: the resulting best-fitting model.  Top right: the best-fitting model, with the masks used in the modeling process.  Bottom left: the normalized residuals.  Bottom middle: the magnification map at the lens plane assuming $z_{\rm s}=1.432$.  \txo{Bottom right: the source plane image, overlaid with the caustic at $z_{\rm s}=1.432$, and labels corresponding to the source galaxies.  
 All panels but the source plane panel (bottom right)} share the same compass and scale bar as the data panel (top left).}
 \label{fig:model}
\end{center}
\end{figure*}

\begin{figure*}
\begin{center}
 \includegraphics[width=1\linewidth]{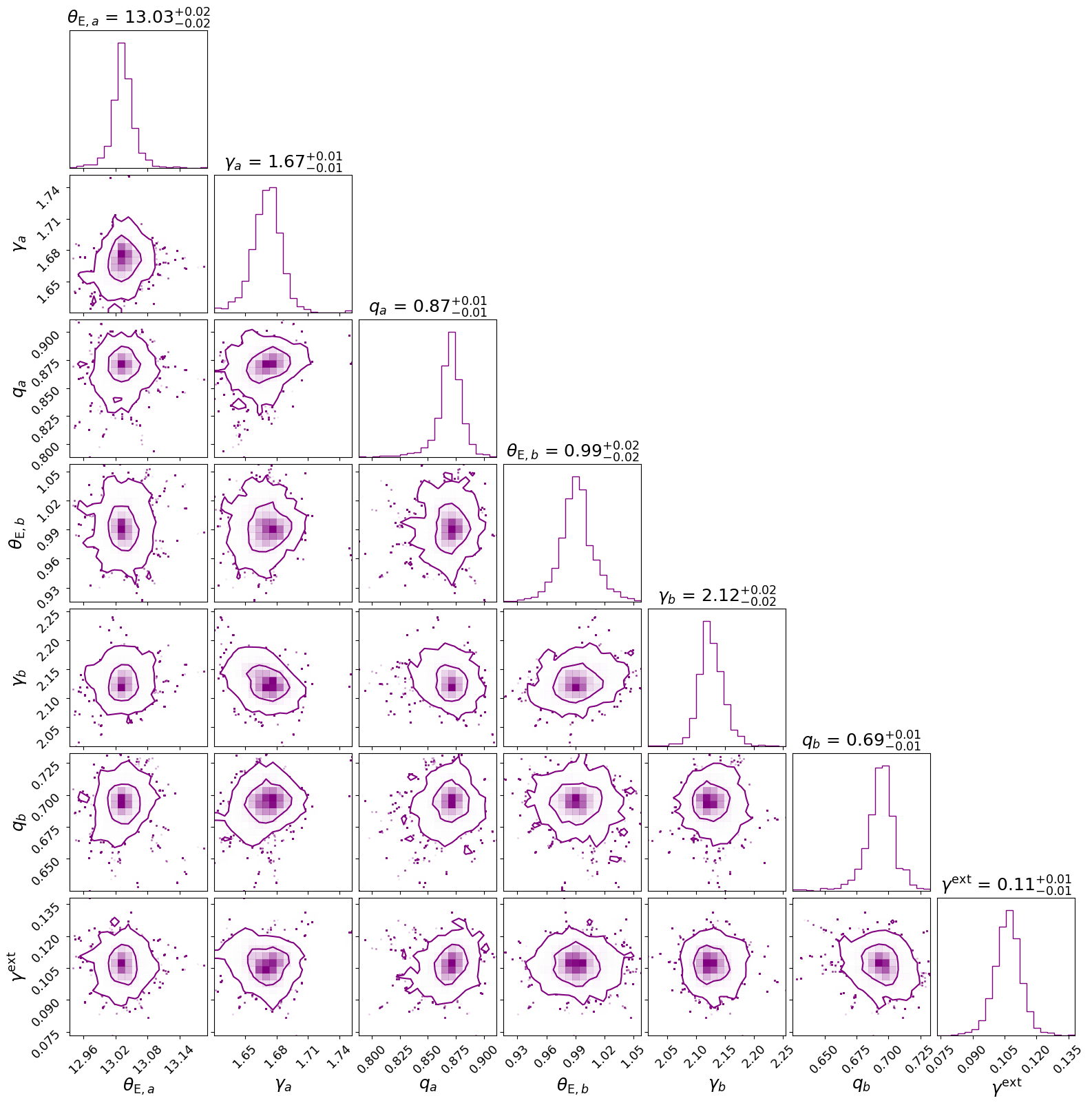}
 \caption{\txr{Posteriors of our lens model parameters.  Also see Table~\ref{tab:lens_model} for a description and summary of each parameter values, as well as the angular configuration components to the lens profiles (i.e., ${\rm PA}_a$, ${\rm PA}_b$, and $\phi^{\rm ext}$).  The inner and outer contours in the 2D distributions represent 68th and 95th percentiles, respectively.}}
 \label{fig:corner}
\end{center}
\end{figure*}

\begin{figure*}
\begin{center}
 \includegraphics[width=5 in]{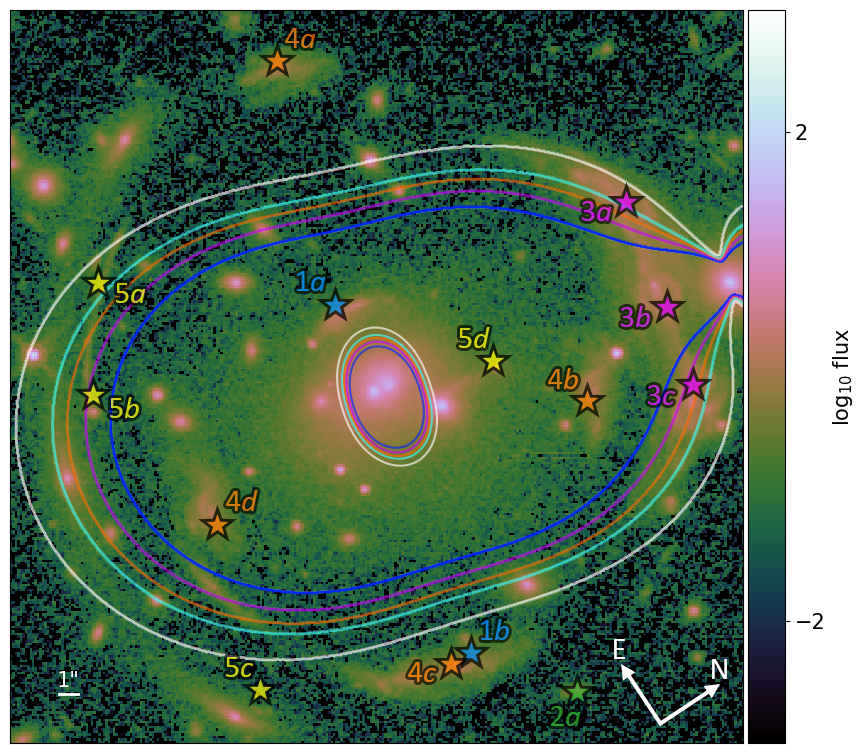}
 \caption{Predicted image locations and critical curves (\txr{blue, purple, orange, cyan, and white for $z_{\rm s}=0.962, 1.166, 1.432, 1.75,$ and $4.52$, respectively}) of our model, plotted over the \textit{HST} F140W image.  We use the centers of the shapelets source light profiles as the source image position on their respective source plane, for source galaxies 1, 3, 4, and 5.  We use the S\'{e}rsic source light profile center for source galaxy 2, as we do not model for an additional shapelets basis.  
 % Shown are the lensed positions of these source galaxies positions.
 }
 \label{fig:clusterplot}
\end{center}
\end{figure*}
We then fit our model to the data, with first a particle swarm optimization operation to locate a maximum (likely close to the global) of the lens likelihood function, then a Markov chain Monte Carlo (MCMC) algorithm with \textsc{emcee} to probe the statistical uncertainties of our model parameters \citep{foreman2013}.  The model and its residuals are shown in Figure~\ref{fig:model}, our parameterizations and statistical uncertainties in Table~\ref{tab:lens_model}\txr{, and the model posteriors are presented in Figure~\ref{fig:corner}.}  
\txr{Additionally, we show our model's predicted lensed image locations over the \textit{HST} F140W image in Figure~\ref{fig:clusterplot}.  We accomplish this by finding the lensed positions of the shapelets source light profile centers
for galaxies 1, 3, 4, and 5 (since the shapelets light contribution exhibits a higher peak flux than their S\'{e}rsic counterpart); we use the S\'{e}rsic profile center for galaxy 2 as we do not model a shapelets profile for it.}

The model achieves a reduced $\chi^2$ of \txr{2.3}. 
For the primary lensing power-law profile, we 
% recover lensing parameterizations of the 
find the best-fit Einstein radius to be \txr{$\theta_{\rm E} = 13.03'' \pm 0.02$} (using $z_{\rm s} = 1.432$) and the power-law slope, \txr{$\gamma = 1.67 \pm 0.01$} (corresponding to a total two-dimensional projected mass within the Einstein radius of \txr{$\sn{4.78}{13} M_\odot$}).  This is significantly steeper ($\sim 3 \sigma$) than population-level total densities of galaxy clusters at $z = 0.2 - 0.3$ \citep{newman2013}, but could be explained by our model systematics (as discussed in the paragraph after next), and/or galaxy cluster density profile evolution \citep{mostoghiu2019}.  

Through the trial-and-error of our modeling process, we made several revelations that helped us in identifying the source images.  We had initially thought image $2a$ to be the counterarc of image $1a$ due to their similar redshifts ($z_s=0.96196$ for $1a$ and $z_s=0.96154$ for $2a$; see Table~\ref{tab:lens_redshifts}), relative locations, and similar colors.  However, from inconsistencies with our predicted counterarc location, we correctly found image $1b$ to be the counter image and $2a$ as a separate lensed galaxy (which is now spectroscopically confirmed; see Table~\ref{tab:lens_redshifts}).  Additionally, through 
constraining images $5a$ and $5b$, our model 
% constrained 
placed source galaxy 5 to lie just inside the caustic, predicting two additional images near $5c$ and $5d$ (commonly referred to as a ``fold'' configuration).  After re-examining the MUSE data, we confirmed $5c$ to indeed be a lensed image, though $5d$ remains to be too dim to resolve its spectra.  %From our best-fit model, we find that lensed images $5c$ and $5d$ that correspond to source galaxy 5, as their predicted images lie near the observed objects $5c$ and $5d$. 

Note that the modeling uncertainties presented in this \txr{paper} are only statistical.
For example, our relatively simple model assumes that the primary lens has a uniform $\gamma$. 
% each lensed source provides an increasingly tighter constraint on the lens's power-law slope.  
From N-body simulation for CDM \citep[e.g.,][]{navarro1996a}, the density profile slope of a cluster-scale halo is expected to vary. 
In this work, we do not account for this possible systematic effect (of a radius-dependent $\gamma$) as this would require a significantly more complex model, which we discuss later.
 Hence we only present the statistical uncertainties associated with our lens model.  
% we do not account for this possible systematic effect.
% However over large cluster scales, we can expect $\gamma$ to vary over radius.
The best-fit $\gamma$ we present can be seen as an average logarithmic power-law slope at different Einstein radii (with $z_{\rm s} = 0.962, 1.166,$ and $1.432$).  
Despite using a relatively simple lens model, we are able to adequately model DESI-090.9854-35.9683.
% , as seen in Figure~\ref{fig:model}.  
% Other cluster systems' models have employed \citep[e.g.,][]{caminha2016, mahler2023}, which typically utilize additional X-ray observations to model the gaseous intracluster medium within clusters.  
% Because 
This is likely due to the fact that DESI-090.9854-35.9683 is relaxed and therefore well-approximated as a power law within the Einstein radii regime. % comparatively small and relaxed, and therefore simpler system.
% and less crowded that other modeled lensed clusters, because this paper being a letter of discovery, and because of the limited amount \textit{HST} observations (only F200LP and F140W), we use our simple double power-law profile to model this system.  
%\cite{hilton2021}, using SZ microwave observations, estimate a total cluster mass of $M_{500c} = 4.023 \times 10^{14} M_\odot$
However, future high-resolution X-ray, additional high resolution imaging, and/or deeper spectroscopic observations may warrant increasing the lens model complexity.

\section{Inferring the Redshift of Source Galaxy 7} \label{sec:redshiftarc}
Though we do not have spectroscopic redshift confirmation of source galaxy 7, we can estimate its redshift by extrapolating our lens model.  In our model, the only quantities that change with source redshift are the Einstein radii, 
% of our lensing potentials: 

\begin{equation}
    \theta_{\rm E} \propto \sqrt{\frac{D_{\rm ls}}{D_{\rm s}}},
\end{equation}
where $D_{\rm ls}$ is the angular diameter distance between the lens and source, and $D_{\rm s}$ is the angular diameter distance to the source.  We can then simply use our best-fit lens model, fix all parameters except the Einstein radii of our two lens potentials (while fixing the ratio between the two), and fit for the source galaxy position that best reproduces the image positions and flux ratios of $7a$, $7b$, and $7c$.  From the best-fitting $\theta_{\rm E}$, we can estimate the source redshift by assuming a standard flat $\Lambda$CDM cosmology of $H_0 = 70$ km s$^{-1}$ Mpc$^{-1}$ and $\Omega_{{\rm M}} = 0.3$.  
\begin{deluxetable}{ccccccccc} \label{tab:mags}
\tablewidth{0pt} 
\tablecaption{AB apparent magnitudes of images $7a$, $7b$, and $7c$ in various bands.  The first three rows are obtained using a $1.4''$ diameter aperture photometry.  The last row uses a larger elliptical aperture photometry ($3.7 \times$ more area) as to capture most of the object's flux; this is only applied onto image 7$b$ as there are neighboring galaxies near 7$a$ and $c$ that would significantly skew the flux measurements.  
%For most of the DECam photometry, the signal is consistent with no flux (due to a much higher (i.e., fainter) limiting magnitude compared to \textit{HST}), and so no upper bound is provided.  
The band's effective wavelength is given below the band name.}
\tablehead{
\colhead{Object} & \colhead{DECam~\ipg} & \colhead{F200LP} & \colhead{DECam~\ipr} & \colhead{DECam~\ipi} & \colhead{DECam~\ipz} & \colhead{VISTA J} & \colhead{F140W} & \colhead{VISTA Ks}\\  & $4730$~\AA & $4970$~\AA & $6420$~\AA & $7840$~\AA & $9260$~\AA & $12500$~\AA & $13920$~\AA & $21500$~\AA} %$\lambda_{\rm e}=$
\startdata
$7a$ & $25.1^{+0.6}_{-0.4}$ & $25.5^{+0.4}_{-0.3}$ & $23.7^{+0.2}_{-0.1}$ & $22.5^{+0.1}_{-0.1}$ & $21.9^{+0.1}_{-0.1}$ & $23.1^{+0.4}_{-0.3}$ & $21.35^{+0.01}_{-0.01}$ & $21.5^{+0.5}_{-0.3}$ \\
$7b$ & $25.2^{+0.7}_{-0.4}$ & $25.1^{+0.2}_{-0.2}$ & $23.1^{+0.1}_{-0.1}$ & $22.3^{+0.1}_{-0.1}$ & $21.6^{+0.1}_{-0.1}$ & $22.2^{+0.1}_{-0.1}$ & $21.15^{+0.01}_{-0.01}$ & $21.1^{+0.3}_{-0.2}$ \\
$7c$ & $24.8^{+0.4}_{-0.3}$ & $25.7^{+0.5}_{-0.3}$ & $23.6^{+0.2}_{-0.1}$ & $22.6^{+0.1}_{-0.1}$ & $21.8^{+0.1}_{-0.1}$ & $22.4^{+0.2}_{-0.2}$ & $21.24^{+0.01}_{-0.01}$ & $21.2^{+0.4}_{-0.3}$ \\ \hline
$7b$ & $24.0^{+0.4}_{-0.3}$ & $24.8^{+0.6}_{-0.4}$ & $22.3^{+0.1}_{-0.1}$ & $21.3^{+0.1}_{-0.1}$ & $20.7^{+0.1}_{-0.1}$ & $21.3^{+0.1}_{-0.1}$ & $20.51^{+0.01}_{-0.01}$ & $20.3^{+0.2}_{-0.1}$ \\
\enddata
\end{deluxetable}
%The \textit{HST} F140W observed fluxes (using $0.7''$ aperture photometry) are $24.66 \pm 1.53\text{, }26.00 \pm 1.58, \text{ and } 23.19 \pm 1.49$ $\mu$Jy for images 5a, 5b, and 5b, respectively

The \textit{HST} F140W observed magnitudes are shown in Table~\ref{tab:mags}.  
The image locations are assumed to have generous uncertainties of $\pm 0.5''$, to account for possible systematic uncertainty within our model.  From these quantities, we fit for the source galaxy position and source redshift that maximizes the log-likelihood of the lensed positions and normalized fluxes of images $7a$, $7b$, and $7c$.

Using this scheme, we estimate source galaxy 7 has a redshift of \txr{$z_{\rm s}=4.52^{+1.03}_{-0.71}$, and magnifications of $8.56^{+0.20}_{-0.19}$, $12.86^{+0.36}_{-0.34}$, $9.73^{+0.26}_{-0.25}$} at $7a$, $7b$, and $7c$, respectively.  While we do not include image $7d$ in our calculation (since its location and flux are not as well defined as the other images), our best-fitting results predict a radial arc within $\sim 1''$ of $7d$.  We reiterate that systematic effects are not accounted for within our models; this estimate assumes a constant $\gamma$ for the modeled perturbers and the uncertainty is purely statistical (see \S\ref{sec:modeling}).  Assuming this redshift, it should be possible to detect Lyman-$\alpha$ emission within the MUSE data, but we do not detect any prominent lines, possibly due to its absence or the low signal-to-noise of source galaxy 7 in the MUSE data.

% Additionally, photometric data from various bands generally agree with our prediction.  From Table~\ref{tab:mags}, we see that on average across the three observed images, the largest magnitude drop lies between DECam~\ipg and DECam~\ipr, which would be where the Lyman-$\alpha$ break lie (assuming $z_{\rm s}=4.65^{+1.02}_{-0.73}$, the Lyman limit would be at $5153^{+930}_{-666} \text{ \AA }$).
One possibility is that the \ion{Mg}{2} emission line doublet ($\lambda \lambda$~2796 and 2803; typically prominent in the broad-line region of AGNs) may contribute to the F140W flux at that redshift, and could partially justify its brightness.  
The general trend in the SED identified from our photometry seem qualitatively similar to the NIRSpec/PRISM spectra of MSAID45924 \citep{greene2023}, a $z=4.46$ AGN identified by the \textit{James Webb Space Telescope} (\textit{JWST}), though it does not have prominent \ion{Mg}{2} emission.  However, given we do not observed a point source in any of the images, we find this conclusion unlikely. %Upon closer inspection, there seems to be possible point-source-like spikes around the center of image 7$b$ (see Figure~\ref{fig:rgb}), though it is difficult to ascertain.
Another possibility is that this is a high-redshift quiescent galaxy, due to its red color across $5000 - 21000$ \AA~(see Table~\ref{tab:mags}).  \textit{HST} has identified similar red, quiescent galaxies up to $z=4$ \citep[e.g.,][]{glazebrook2017, schreiber2018}, and \textit{JWST} up to $z=5$ \citep[e.g.,][]{carnall2023, setton2024, degraaff2024}.  This theory is supported by the fact that while we faintly observe a continuum for source 7 in the MUSE data, we do not observe any identifiable emission lines, which could indicate quiescence.
If this is the case, source galaxy 7 could serve as additional evidence for early galaxy quenching more efficiently than previously understood.  With deeper spectroscopy, we will be able to determine the identity and redshift of source galaxy 7 with more confidence.
% While these pieces of evidence are circumstantial, as it is based on ground-based photometry with generous uncertainties, it nevertheless suggests that our prediction may be correct.

\txr{We do not apply the same analysis on source 6 as we did for source 7 for two reasons.  Firstly, we are not certain that the images $6a$ and $6b$ are indeed of the same source galaxy.  Unlike with source 7, where it is visually obvious that $7a$, $b$, and $c$ belong to the same source (due to striking similarity in colors and a textbook orientation of its lensed images), this is not the case for source 6.  Secondly, images $6a$ and $6b$ are much more diffused, complex, and crowded (hence much more difficult to accurately recover photometry for) compared to the compact images of $7a$, $b$, and $c$.  As the photometry is a necessary part of the analysis performed on source 7, this would make it much more difficult to accurately estimate a redshift for source 6.  Instead, if we assume that images $6a$ and $6b$ are the same source, we can deduce that the outer critical curve should cross between the two images.  From this, our model infers that the redshift of source 6 should lie between $1.75 \le z_{\rm s} \le 4.52$ (see the cyan and white contours of Figure~\ref{fig:clusterplot}).}

\section{Discussion and Conclusion} \label{sec:conclusion}
In this \txr{paper}, we present DESI-090.9854-35.9683, a cluster-scale lens first identified in \citet{jacobs2019}, and later independently found in \cite{huang2021} and \cite{odonnell2022a}.  From \textit{HST} observations in the F140W and F200LP bands (16773; K. Glazebrook), we identify seven possible lensed sources.  With spatially resolved spectroscopic observations from MUSE, we confirm five of these lensed sources, 
% (while the other two remain uncertain), 
and provide the proper velocities for the resolved images.  From this, we construct a simple yet successful
lens model using only two elliptical power law mass profiles and external shear. 
% to well-recover the observed F140W image, 
\txr{We use this model to infer that one of the spectroscopically unresolved sources (source galaxy 7) is at a redshift of \txr{$z_{\rm s}=4.52^{+1.03}_{-0.71}$}, while the other (source galaxy 6) is at a redshift range of $1.75 \le z_{\rm s} \le 4.52$.  }

In this spectacular system, nearly all the ``classic'' lensing configurations are represented. Arc family 2 is singly imaged, \txr{as its source is outside of the caustics. The source for arc family 1 is well inside the radial caustic but just outside the tangential (diamond) caustic, and thus is doubly lensed (though there ought to be a third, highly de-magnified image near the lens center).}
Arc family 5 has a predicted fourth image matching an observed object, and forms a ``fold" configuration.  Whereas arc family 3 has three images, exhibiting a ``naked cusp'' configuration.  Arc family 4 forms a quintessential Einstein cross.
Finally, arc family 7 appears to have four images, including a possible radial arc, and if confirmed, this would form a ``cusp'' configuration \citep[likely, also a naked cusp; see Figure~2 of][]{lewis2002}.

With additional spectroscopic observations to confirm source galaxies 6 and 7, deeper X-ray observations to better constrain the gaseous intracluster medium of this system \citep[e.g.,][]{caminha2016, mahler2023}, and/or deeper higher-resolution imaging, we can construct an even more accurate model of this system with fewer assumptions of the total mass profile.  
% By studying and well-modeling a lens with multiple source galaxy, 
In addition to the potential systemic effect of $\gamma$ varying with cluster-centric radius (see \S\ref{sec:modeling}), 
another possible systematic effect that a more comprehensive lens model may need to take into account is the gravitational effects of the nearby sources on sources that are further away  (i.e., not just multiple source planes, but also multiple lens planes).
It then becomes possible to constrain cosmological parameters such as $\Omega_{{\rm M}}$ and $w$ \citep[e.g.,][]{collett2014}.  
Furthermore, the large number of lensed sources can translate to tighter constraints on the matter profile of the lens galaxy, allowing for closer examination of the dark matter and luminous matter distributions/interactions in galaxy clusters.  

Clearly, there is much more to be learned from this system.
% , of which we have only begun the scratch the surface.  
The coincidental alignment of seven galaxies and a foreground galaxy cluster can give us unprecedented insight into the universe, whether it be high redshift galaxies, cluster properties, or cosmology.  

% In the case of strong lenses with multiple source galaxies at different redshifts, 
% it is possible to constrain cosmological parameters such as $\Omega_{{\rm M}}$ and  the dark energy equation of state, $w$ \citep[e.g.,][]{collett2014}.  
% % This is done by modeling for mass profile(s) at the lower-redshift source galaxy/galaxies, and fitting for the relative lensing strength of the primary deflector to the different source galaxies ($\beta$), as only a function of $w$, $\Omega_M$, and $\Omega_k$ in a flat $w$CDM universe.  
% While we do not perform this analysis in this letter, future analysis for DESI-090.9854-35.9683 may be particularly insightful given the number of lensed sources and the perceived simplicity/symmetry of the primary lens. 

\section*{Acknowledgments} \label{sec:acknowledgments}
This work was supported in part by the Director, Office of Science, Office of High Energy Physics of the US
Department of Energy under contract No. DE-AC025CH11231. This research used resources of the National Energy
Research Scientific Computing Center (NERSC), a U.S. Department of Energy Office of Science User Facility operated
under the same contract as above and the Computational HEP program in The Department of Energy’s Science Office
of High Energy Physics provided resources through the “Cosmology Data Repository” project (Grant \#KA2401022).  This research was supported by the Australian Research Council Centre of Excellence for All Sky Astrophysics in 3 Dimensions (ASTRO 3D), through project number CE170100013.  
The work of A.C. is supported by NOIRLab, which is managed by the Association of Universities for Research in Astronomy (AURA) under a cooperative agreement with the National Science Foundation. 
X.H. acknowledges the University of San Francisco Faculty Development Fund.
T.J. and K.V.G.C. gratefully acknowledge financial support from the National Science Foundation through grant AST-2108515, NASA through grant HST-GO-16773, the Gordon and Betty Moore Foundation through Grant GBMF8549, and from a Dean’s Faculty Fellowship.  We thank Ned Taylor and Michelle Cluver of the Swinburne University of Technology for providing reduced VISTA 4MOST Hemisphere Survey data.

\software{Lenstronomy \citep{birrer2018, lenstronomyII},
          NumPy \citep{numpy}, 
          SciPy \citep{scipy},
          Matplotlib \citep{matplotlib}, 
          Astropy \citep{astropy2013, astropy2018, astropy2022}, 
          Emcee \citep{foreman2013}, 
          Jupyter \citep{kluyver2016},
          ESO Recipe Execution Tool \citep{2020A&A...641A..28W},
          Zurich Atmosphere Purge Sky Extraction Tool \citep{2016MNRAS.458.3210S}.
          }

\bibliography{sample631}{}
\bibliographystyle{aasjournal}
\end{document}